\documentclass[onecolumn,natbib,lscape]{aa}
\usepackage{natbib}
\usepackage{graphicx}
\usepackage{graphics}
\usepackage{amssymb}
\def\arcmin{\hbox{$^\prime$}}
\def\deg{\hbox{$^\circ$}}
\def\rxa{RXJ\,1821.6$+$6827~}

\def\rxb{RXJ\,1716.6$+$6708~}
\def\ms{MS\,1054$-$0321~}

\begin{document}
   \title{RXJ\,1821.6$+$6827: a Cool Cluster at z$=$0.81 from the {\em ROSAT\/}
                              NEP Survey
      \thanks{Data presented here used the XMM-{\it Newton} facility}
}
   \author{I. M. Gioia          
          \inst{1,6,7},
          A. Wolter
          \inst{2},
          C. R. Mullis          
          \inst{3,6,7},
          J. P. Henry
          \inst{4,7} 
          H. B\"ohringer
          \inst{5} and
          U. G. Briel
          \inst{5}
          }

   \offprints{I. M. Gioia}

  \institute{Istituto di Radioastronomia CNR, Via P. Gobetti 1
             I-40129 Bologna, Italy \\
             \email{gioia@ira.cnr.it}
         \and
            INAF-Osservatorio Astronomico di Brera, Via Brera 28,
             I-20121 Milano, Italy \\
             \email{anna@brera.mi.astro.it}
          \and
          European Southern Observatory, Karl-Schwarzschild-Str. 2, 
          Garching bei M\"unchen, D-85740 Germany \\
             \email{cmullis@eso.org}
         \and
            Institute for Astronomy, 2680 Woodlawn Drive,
            Honolulu, HI 96822, USA \\
             \email{henry@ifa.hawaii.edu}
          \and
          Max-Planck-Institut fur Extraterrestrische Physik,
         Giessenbachstrasse, Postfach 1312, Garching, \\ 
                    D-85741 Germany \\
          \email{hxb@xray.mpe.mpg.de,ugb@xray.mpe.mpg.de}
          \and
	  Visiting Astronomer at the W. M. Keck Observatory, jointly
	  operated by the  California Institute of Technology, the
	  University of California and the National Aeronautics and Space
	  Administration.
	  \and 
	  Visiting Astronomer at the Canada-France-Hawai$'$i Telescope,
	  operated by the National Research Council of Canada, le Centre
	  National de la Recherche Scientifique de France and the University
	  of Hawai$'$i.
           }

   \date{Received ........; accepted .............}

 \abstract{
We present an analysis of the properties of the cluster of galaxies 
RXJ\,1821.6$+$6827, or NEP\,5281, at a redshift z$=$0.816$\pm$0.001.
RXJ\,1821.6$+$6827 was discovered  during the optical identification 
of the X-ray sources  in the North  Ecliptic Pole (NEP) region of the 
{\it ROSAT} All-Sky Survey and it is the  highest  redshift cluster 
of galaxies of the NEP  survey. We have measured spectroscopic 
redshifts for twenty cluster galaxies  using the Keck-I and the 
Canada-France-Hawai$'$i (CFH) telescopes.  The value for the cluster  
velocity dispersion is  $\sigma_{V}=775^{+182}_{-113}$ km s$^{-1}$. 
The cluster was also observed   by XMM-{\em Newton}. Both the optical and 
X-ray data are presented in this  paper. The cluster has an unabsorbed 
X-ray flux in the 2--10 keV energy band of 
$F_{2-10 keV} = 1.24^{+0.16}_{-0.23} \times10^{-13}$ erg cm$^{-2}$ s$^{-1}$
and  a K-corrected luminosity in the same band of
$L_{2-10 keV} = 6.32_{-0.73}^{+0.76} \times10^{44}$ h$^{-2}_{50}$  erg s$^{-1}$
(90\%  confidence level). The cluster X-ray bolometric luminosity is
$L_{BOL,X} = 1.35^{+0.08}_{-0.21} \times10^{45}$ h$_{50}^{-2}$ erg s$^{-1}$
($L_{BOL,X} = 1.17^{+0.13}_{-0.18} \times10^{45}$ h$_{70}^{-2}$ erg s$^{-1}$
in the concordance cosmology). The data do not allow fitting both metal 
abundance  and temperature at the same time. 
The abundance is unconstrained  and can  vary in the range 
0.28--1.42 Z$_{\sun}$ while the best fit X-ray 
temperature is T$=$4.7$^{+1.2}_{-0.7}$ keV. This emission weighted X-ray 
temperature is a little lower, barely within the uncertainties, 
than the  predicted   temperature, T$=$6.34$^{+0.13}_{-0.35}$ keV, from the 
L$_{X}-T_{X}$ relation of local  clusters published in the literature.  
The optically measured velocity  dispersion is  consistent with the velocity 
dispersion expected from  the   $\sigma_{V}-T_{X}$ relationship.  
We also examine the point X-ray source RXJ\,1821.9$+$6818, or NEP\,5330,
located to the south  east of the cluster which was identified as a QSO 
at z$=$1.692$\pm$0.008 in the {\it ROSAT} NEP survey. The X-ray source is well 
fitted by an absorbed power law model  with 
N$_{H}=4.1_{-3.0}^{+3.1}\times10^{20}$ atoms cm$^{-2}$  and a photon index 
$\Gamma=1.67\pm$0.12 typical of an active galactic nucleus.

}

\authorrunning{Gioia et al.}
\titlerunning{RXJ\,1821.6$+$6827: a cool cluster}

\maketitle

%

\section{Introduction}

Clusters of galaxies are the most massive, collapsed structures in the 
Universe. They are the highest peaks in a cosmic terrain driven by 
gravitational clustering and represent the manifestations of 
cosmic structure building \citep{peeb93, pea99}.  The study of 
the properties of clusters at intermediate and high redshift, 
both individually and as a population, is important since it allows
different cosmological and structure formation models to be tested and 
constrained. The internal mix of components within clusters, as  well as the 
space density of the most  distant and massive clusters and the 
temperature distribution function  of X-ray  clusters, can be used to 
determine fundamental cosmological  parameters  \citep{ha91, ob92, ob97, 
eke98, bor99,  bor01, hen00, hen04, vik03}.
Through the detection  of X-ray emission from the hot intracluster 
gas, X-ray cluster surveys are unbiased in the sense that they exclusively  
select gravitationally bound  objects and are essentially unaffected by  
projection effects. There are only a few known clusters of galaxies beyond 
redshift unity, most of them were found in X-ray survey samples 
\citep[see][]{ros04}. 
Even if we are far from having very large complete samples 
of distant ($z>0.8$) clusters, our capabilities to discover and observe 
distant objects are rapidly increasing as more accurate and sensitive 
observing techniques become available. 

In recent years we have been involved  in the study of the deepest region 
of the {\it ROSAT} All-Sky Survey, at the North Ecliptic Pole \citep{hen01, 
vog01}, to produce a complete and unbiased X-ray-selected  
sample of clusters of galaxies.  The resulting NEP cluster 
sample has been used to investigate the nature of cluster evolution 
\citep{gio01} and to explore the potential implications for large-scale 
structure models \citep{mul01}. RXJ\,1821.6$+$6827, or   NEP\,5281,
was detectd in  the  {\it ROSAT} NEP  survey as a 4.5$\sigma$ source with only 
40$\pm$9  net photons in a vignetting corrected exposure of 5519s. 
The identification program of the survey \citep{gio03} revealed the 
cluster nature of the source. Optical spectroscopic observations with 
the CFH and Keck-I telescopes placed  the 
cluster at a redshift of z$=$0.816$\pm$0.001. \rxb is  the most distant 
cluster of galaxies in the {\it ROSAT} NEP survey. Follow-up XMM-{\em Newton}
observations were also performed, and are reported here together with 
the optical observations. Note that this is the second z$=$0.81 cluster 
discovered in the NEP survey. The first one, RXJ\,1716.6$+$6708 \citep{gio99},
does not  appear in the final complete NEP sample because it did not meet 
the  selection criterium for source count rate of signal-to-noise ratio 
$\geq 4\sigma$.

Section 2 describes the optical and X-ray data acquisition and 
analysis. Section 3 details the results obtained and discuss these results. 
A brief summary is given in Section 4. 
Throughout this paper, we assume an Einstein-de Sitter 
model  H$_{0}=50$ km s$^{-1}$ Mpc$^{-1}$,  $\Omega_M=1$, and 
$\Omega_\Lambda=0$ for direct comparison to previous work
in this field, but we also repeat the calculations in the  current
cosmological concordance model ($h, \Omega_M, \Omega_\Lambda$)=(0.7, 0.3, 
0.7). At the redshift of the cluster, the luminosity distance is 5.62 
h$_{50}^{-1}$ Gpc, the angular size distance is 1.70 h$_{50}^{-1}$ Gpc,
and the scale is 8.26 h$_{50}^{-1}$ kpc per arcsec.
Throughout the paper quoted  uncertainties are  90\% confidence levels for 
one  interesting parameter.


\section{Observations and Data Analysis}

In this section we present the optical spectroscopy for the cluster 
galaxies performed at the  CFHT and  Keck-I and the  X-ray follow-up 
observations  of RXJ\,1821.6$+$6827 acquired  with 
XMM-{\em Newton}.  We first  describe  the optical spectroscopy, 
and then present the X-ray data which show that NEP\,5281 has a low
temperature, lower than expected from its bolometric luminosity, but 
commensurate with its velocity dispersion. We also present the
optical and X-ray data on the point source to the south east of the 
cluster, namely RXJ\,1821.9$+$6818, or NEP\,5330, which is identified 
with a QSO at z=1.692.

   \begin{figure}
   \centering
   \includegraphics[width=15cm]{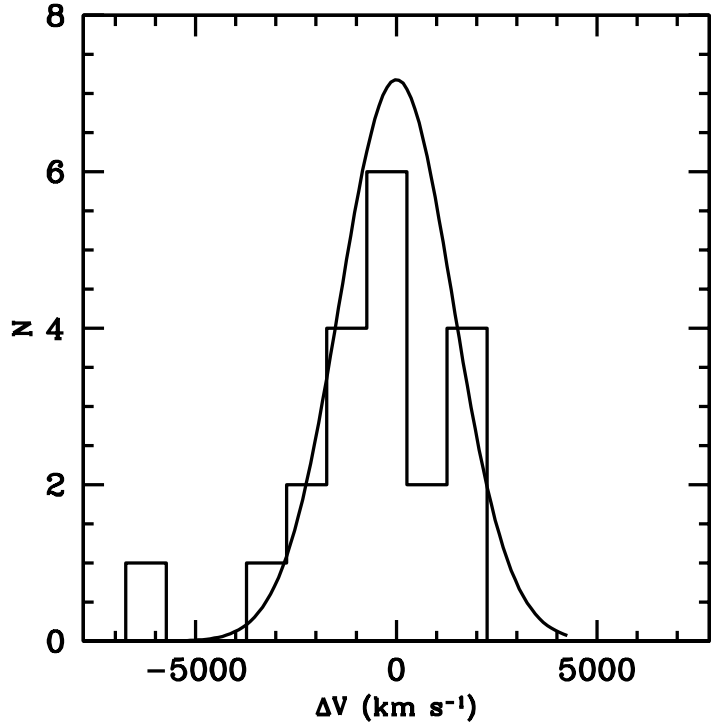}
      \caption{Distribution of velocities for the 20 galaxies showing a low
velocity tail at $\Delta_v\sim$ $-$6000 km s$^{-1}$.  The curve shows a 
Gaussian distribution centered at $<v>=244,736$ km s$^{-1}$ with width given 
by the derived  $\sigma_{V}=1407$ km s$^{-1}$ (not corrected to the first 
order for cosmological redshift) and normalized for the 18 galaxies
considered as cluster members.}
         \label{Fig1}
   \end{figure}

\subsection{CFHT and Keck-I Spectroscopy}

RXJ\,1821.6$+$6827 was observed on July 3, 1997, at the CFHT
with the MOS instrument using the STIS2 2048$^{2}$  and the O300 
grism in multislit mode. The wavelength coverage was from 4000 to 10000 \AA\
and the pixel size was 4.8 \AA\ pix$^{-1}$. The slitlet width of
1.5$''$  provided a spectral resolution of 17 \AA. The data were reduced
using the MULTIRED package developed by \cite{olf95}.
Seven galaxies from the CFHT data  were identified as cluster members 
from the CaII break only (one has [OII] in emission). 
NEP\,5281 was also observed at the Keck-I telescope on June 22, 2001, 
with the  Low Resolution and  Imaging Spectrograph \citep[LRIS,][]{oke95}
in  slit-mask mode.  With both spectrographs in operation, 
LRIS\_R and LRIS\_B,  it is possible to obtain multi-object spectra 
covering the entire optical window in one integration. However we used 
only data from the red  arm given the redshift of the cluster. The 600 
l/mm grating blazed at 7850 \AA, to target [OII] plus Ca II break and CN
at the redshift of the cluster. Combined  with our grating angle the
wavelength coverage was  approximately 6500 \AA\ to 9500 \AA, and the pixel 
size was 1.25 \AA.  The GG495 filter was used to eliminate the overlapping 
second order spectrum to avoid any contamination blueward of
9500 \AA.  Given the fact that galaxies are extended objects, even at 
this redshift, we used a slit width of 1.4\arcsec, which gives
a spectral resolution of roughly 6.5 \AA. For the selection of the objects 
to spectroscopically observe, we  used  deep two color (B and I)
images previously taken by us at the University of Hawai$'$i (UH) 2.2m 
telescope. The UCSCLRIS package designed by Drew Phillips and collaborators 
at Lick Observatory was used to prepare the slit-mask files.  We designed
two masks at different position angles in order to cover
the central  part of the cluster. Only data from the first slitmask were
used since  the second slitmask exposure was affected by heavy cirrus 
and bad weather. Thus we do not have a complete coverage for the galaxies 
in the central region of the cluster. The first mask was exposed for 
10,800s. After the mask exposure, flat fields and arc calibrations exposures 
were also obtained. The data have been reduced using the standard IRAF 
package routines for 2-D spectra images. The two-dimensional spectra had 
to be straightned  using the software package WMKOLRIS kindly provided by  
Greg Wirth at the W. M. Keck Observatory.

Fifteen galaxies out of 26 observed objects were recognized as cluster 
members on the basis of CaII H and K, CN and, in some cases, the G band 
absorption features. Only one of the 15 galaxies observed had [OII] in 
emission. Two of the seven galaxies observed at the CFHT are common to the 
Keck-I observations.  In the end, 15 unique objects observed by Keck plus 
5 unique objects observed by CFHT turned  out to be cluster  member galaxies. 
These twenty  galaxies  are listed in Table 1. 
For  each galaxy the (J2000) coordinates, measured velocity, 1$\sigma$ 
error, and redshift are given. In the last column the
main spectral features are also noted.

The velocity  histogram for the 20 member galaxies is shown in Fig.~\ref{Fig1}.
There is a low-velocity extension in the histogram at 238,500 km s$^{-1}$ 
(2 galaxies)  that our 3$\sigma$ iterative clipping algorithm 
\citep[following][]{dan80} excludes from the computation of the cluster 
velocity.  From 18 accepted cluster members, and taking into account the 
errors on the redshift, we obtain a mean velocity  $<v>=244,736\pm325$  
km s$^{-1}$, an average redshift  $<z>=0.8163\pm0.0011$, and a dispersion 
along the line of sight  $\sigma_{V}=775^{+182}_{-113}$  km s$^{-1}$ 
(corrected to first  order for cosmological redshift by dividing by 1$+$z). 
Without the 3$\sigma$ clipping the mean velocity for the 20 galaxies
is  $<v>=244,249\pm384$  km s$^{-1}$, the average redshift is  
$<z>=0.8148\pm0.0016$, and the dispersion along the line 
of sight is  $\sigma_{V}=1142^{+247}_{-155}$  km s$^{-1}$.

  \begin{figure}
   \centering
\caption{Figure 2 can be viewed and downloaded from this http URL
          http://www.ira.cnr.it/~gioia/PUB/publications.html.
            The figure shows 
            a 1800s I-band image of \rxa taken at the
            UH 2.2m telescope.  The combined MOS1 and  MOS2 
            image was smoothed with a Gaussian to have 16 pixels
            per cell (corresponding to 8\arcsec). The energy range is
            0.3--8 keV band. Units for the smoothed overlayed X-ray contours 
            are (6.5, 7.8, 9.1, 13, 18.2, 24.7, 32.5) 10$^{-3}$ counts 
            s$^{-1}$ arcmin$^{-2}$ over the background which correspond 
            to (2.5, 3, 3.5, 5, 7, 9.5, 12.5)$\sigma$ over the 
            background.  The circles indicate 14 out of 20 cluster members. 
            The cD galaxy is marked as galaxy \# 7. The
            other 6 galaxies  at the redshift of the cluster fall 
            outside the field of view  (see Table 1). The image measures 
            $1092 \times 1092$ pixels covering a field  of $4'\times4'$ 
            ($1.98h_{50}^{-1}\times1.98 h_{50}^{-1}$ Mpc at $z =$ 0.8163).}
         \label{Fig2}
   \end{figure}

Using the ROSTAT software by Tim Beers  and collaborators
\citep{beers90} \citep[see also][]{bb93} we obtain very  similar results
to those of the  3$\sigma$ iterative clipping algorithm by \cite{dan80}.
Below the values obtained by the biweight estimator of scale, as suggested by 
\cite{beers90} for this number of galaxies, are given. 
The errors are the  90\% confidence intervals. 
The average redshift is $<z>=0.8156\pm0.0024$ 
accounting for errors on z ($<z>=0.8147\pm0.0044$ with no errors), 
and a $\sigma_{V}=577\pm231$ km s$^{-1}$, corrected to first order for 
cosmological redshift ($\sigma_{V}=1048\pm421$ km s$^{-1}$ corrected to 
first order cosmological redshift but without taking into account the errors 
on redshift),  which are consistent with  either the 20 cluster member and 
the  18 cluster member velocity  dispersions obtained with the previously 
adopted  3$\sigma$ clipping algorithm.

  \begin{figure}
   \centering
   \includegraphics[width=12cm, angle=-90]{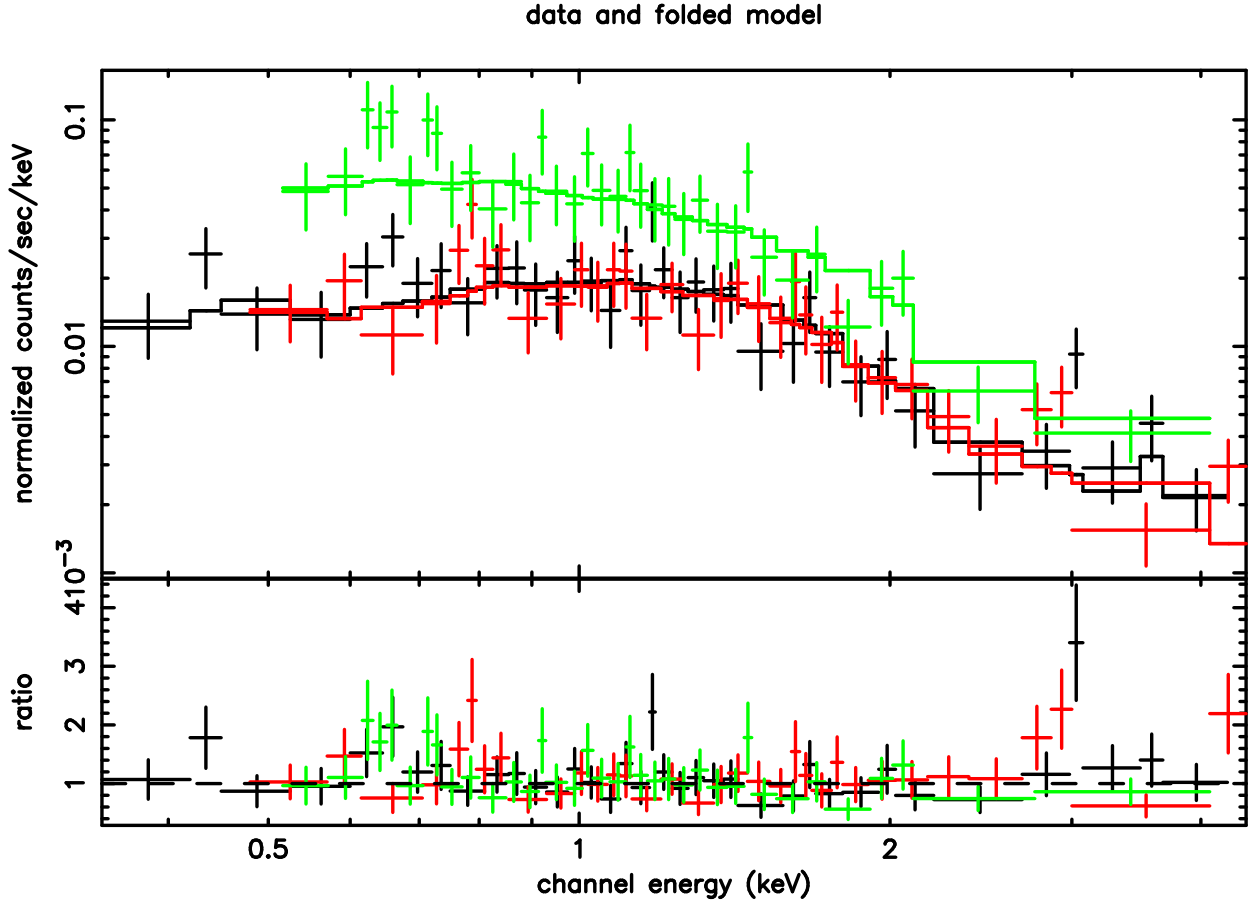}
     \caption{Binned X-ray spectrum and residuals for \rxa.
         The spectrum was binned in such a way that each  resulting 
         energy channel has a signal-to-noise ratio of at least 4.
         The solid line represents the best fit MEKAL model. The top 
         grey spectrum is the pn spectrum, while the lower and darker
         spectra are the MOS1 and MOS2 spectra.} 
        \label{Fig3}
   \end{figure}

\subsection{XMM-{\em Newton} observations}

XMM-{\it Newton} \citep{jan01} observed \rxa as part of the GO program
in three epochs, 
on November 13 and 23, 2002, and  on December 19, 2002,  for a nominal 
exposure of 63 Ks with the  European Photon Imaging Camera (EPIC) pn 
\citep{sru01} and for 73.5 ks with the EPIC  MOS CCD arrays \citep{tur01}.
The pn was operating  in extended-full-frame mode with thin filter while 
the MOS was in the full-frame mode also with the thin filter applied. 
The pileup is not a problem given the low count rate of the X-ray source.

Unfortunately the observations suffer from periods of
very high background. Event files produced by the standard pipeline  
processing have been examined and filtered to remove the high background time 
intervals (using the version 5.4.1 of the Science Analysis Software, SAS, 
and  the latest calibration files released by the EPIC team). Only events 
corresponding to pattern 0-12 for MOS and pattern 0-4 for pn have been 
used\footnote{see the XMM-{\it Newton} Users' Handbook \\
{\it http://xmm.vilspa.esa.es/external/xmm\_user\_support/documentation/uhb/XMM\_UHB.htm}}.
Good time intervals were selected by applying
thresholds of 0.35 counts s$^{-1}$ in the MOS and 1 counts s$^{-1}$ in the 
pn to the photons at energies greater than 10 keV. At these higher energies
counts come mostly from background. 
The net exposure times, after data cleaning, are 21.3 Ks and 22.0 Ks for 
the MOS 1 and MOS 2, respectively, and 10.6 Ks for the pn. For subsequent
analysis background counts have been accumulated using nearby source-free 
circular  regions.

The three images, one for each instrument, taken on different epochs
were summed using the SAS task 
{\em merge}. Given the position of the source at the center of the field, 
the different orientations of the instrument with respect to the sky 
coordinates should not affect the response of the instruments. 
Thus resulting response  matrices and auxiliary files are the average of 
the three exposures, weighted by the different exposure times. 

Response matrices (that include the correction for the effective area) have 
been generated using the SAS tasks {\it arfgen} and {\it rmfgen}. The X-ray 
fluxes reported below are computed using the MOS1 detector and calibration(s). 
From our data the normalization of the MOS2  is  5\% lower than  that of the 
MOS1 whereas the normalization of the pn is 24\% lower. This apparently  
discrepant normalization value has to be attributed to the fact that 
the cluster falls on the CCD gap of the pn.

   \begin{figure}
   \centering
    \includegraphics[width=14cm, angle=0]{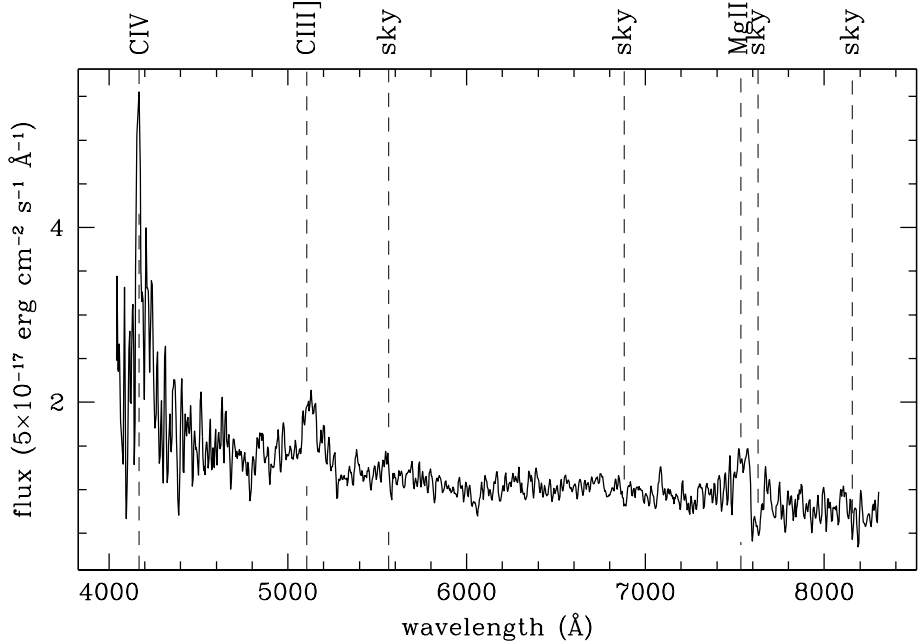}
      \caption{Long-slit spectrum taken at the UH 2.2m of the QSO 
       at z$=$1.692 associated to RXJ\,1821.9$+$6818.
       The total integration time was 
       of 2400s. The dashed lines indicate the positions of the CIV, 
       CIII], Mg emission lines at the QSO redshift. Wavelengths of
       atmospheric absorption bands are also indicated.}
       \label{Fig4}
   \end{figure}

   \begin{figure}
   \centering
    \includegraphics[width=12cm, angle=-90]{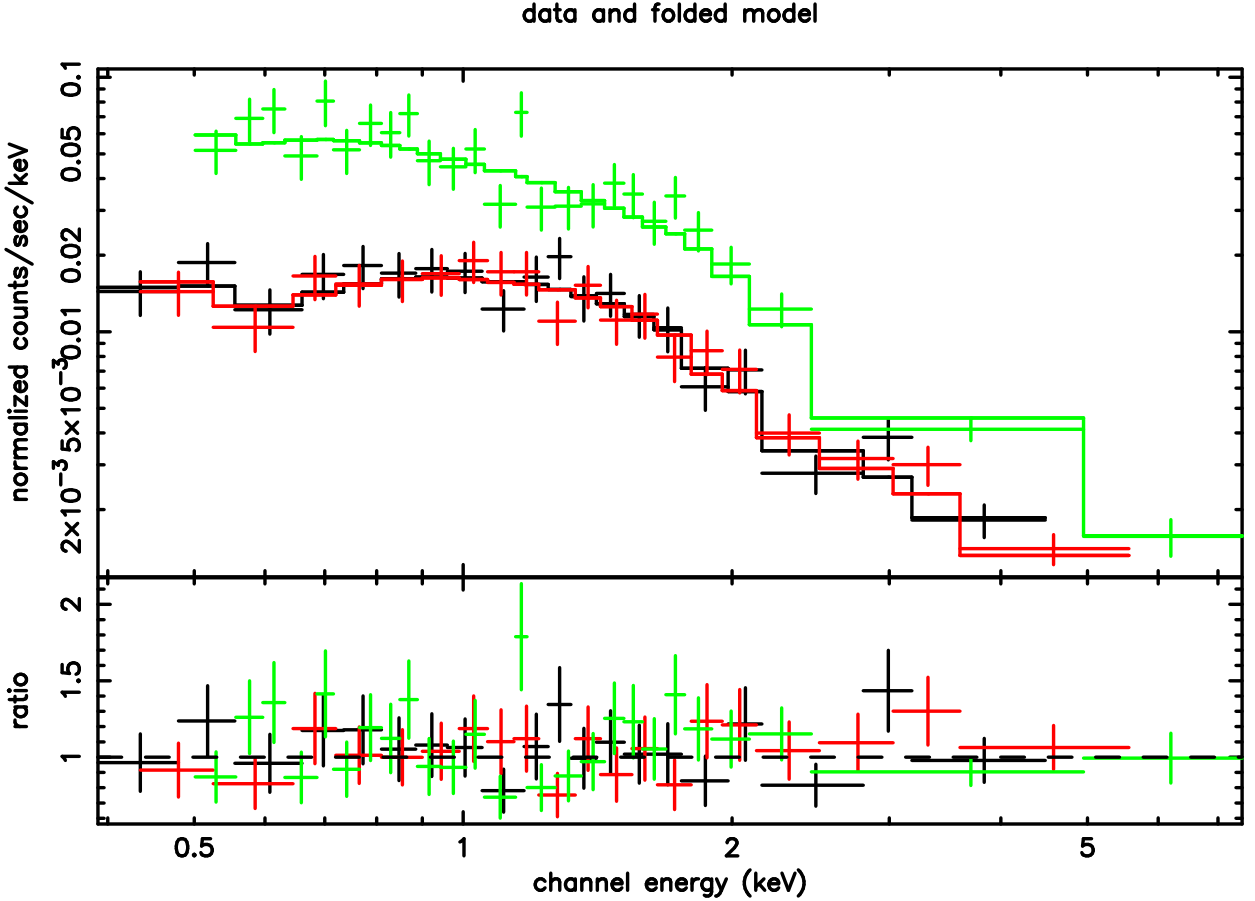}
      \caption{Binned X-ray spectrum and residuals for
         RXJ\,1821.9$+$6818, the QSO in the XMM-{\it Newton} field. 
         The spectrum was binned in such a way that each
         resulting energy channel has a signal-to-noise ratio of at least 5.
         The solid line represents the best fit power law model. The top
         grey spectrum is the pn spectrum, the MOS1 and MOS2 spectra are
         the two lower and darker spectra.}
       \label{Fig5}
   \end{figure}

\subsection{Overall Cluster Properties}

Fig.~\ref{Fig2} shows the XMM-{\it Newton} smoothed contours of
RXJ\,1821.6$+$6827 overlayed onto an optical 1800s I band image taken at the
UH 2.2m telescope. Both MOS1 and MOS2 data were used to produce the 
X-ray contours in the 0.3--8 keV band. The data were smoothed with a Gaussian
using cells of 16  pixels corresponding to a resolution of 8\arcsec.
The X-ray cluster is extended along the NE-SW  direction. There are not 
enough X-ray counts to produce a  more meaningful comparison between the 
optical and X-ray images.

To measure the emission-weighted  cluster temperature, net counts were 
extracted from a circle of radius 100\arcsec ~centered at 
$\alpha_{2000}= 18^{h}~21^{m}~32.5^{s},  ~\delta_{2000} = +68\deg~27\arcmin
~45.0\arcsec$ which corresponds to the barycenter of the X-ray emission
and is very close to the position of galaxy 7 (see Table 1). 
At the redshift of the cluster  such a radius corresponds to 
a linear size of 826 h$_{50}^{-1}$ kpc. A larger radius including more 
extended emission counts could not be chosen given the high increment  in 
background  counts. 
There are 708$\pm$35 net counts in the (0.4--4.0 keV) energy range in the
MOS1 detector, 671$\pm$36 in (0.5--4.3 keV) in the  MOS2, and 751$\pm$40 
net counts in  (0.5--3.4 keV) in the pn  detector. X-ray properties are 
listed given in Table 2.
The data were fitted in XSPEC (version 11.3.0) with a single temperature
MEKAL model  \citep{mew85} modified by Galactic absorption, where the ratio
between the elements are fixed to the solar values as in \cite{ag89}\footnote{
These values for the solar metallicity have recently been superseded by the 
new values of \cite{gs98}, who used a 0.676 times lower Fe solar abundance. 
However, we prefer to report metallicities in the units of Anders \& Grevesse 
abundances since most of the literature still refers to these old values.}.

The relative normalization of the three instruments was left free 
to account for possible intercalibration  differences or  mismatches. 
The absorption and redshift were kept constant while the temperature and
abundance were left free to vary.
The absorption was fixed at the Galactic value of  $5.24\times10^{20}$ 
atoms cm$^{-2}$ \citep{elv94} and the redshift was fixed at z$=$0.8163.
The best fit with a thermal MEKAL model gives a temperature 
T$=$4.7$^{+1.2}_{-0.7}$ keV and a best fit abundance Z$=$0.77 Z$_{\sun}$, 
which  is unconstrained by these data due to a non-detection of the iron 
K line. The metallicity can vary in the range 
0.28--1.42 Z$_{\sun}$. The binned X-ray spectrum and best fit 
folded model are shown in Fig.~\ref{Fig3} where  bins were chosen so as 
to have each resulting energy channel significant at the signal-to-noise  
ratio of at  least 4. The total number of photons in each bin allows the use 
of the  $\chi^{2}$ statistics for the fit. The fit has a reduced $\chi^{2}$ of 
0.90 for 102 degrees of freedom. If one fixes the abundance to 
Z$=$0.5 Z$_{\sun}$
then the  temperature increases  a little (T$=$4.9$^{+1.2}_{-0.8}$ kev) 
but it is consistent with the previously determined temperature within the 
errors. The reduced $\chi^{2}$ becomes  0.90 for 103 degrees of freedom.

The unabsorbed flux in the 2--10 keV energy band is 
$F_{2-10 keV} = 1.24^{+0.16}_{-0.23} \times10^{-13}$ erg cm$^{-2}$ s$^{-1}$,
(see Table 3). While the unabsorbed X-ray flux in 0.5--2.0 keV band is 
$F_{0.5-2.0 keV} = 1.35^{+0.13}_{-0.21} \times10^{-13}$ erg cm$^{-2}$ 
s$^{-1}$, to be compared with the flux obtained with the {\it  ROSAT} 
PSPC in 0.5--2.0 keV of $F_{X,{\it ROSAT}} = 1.02 \pm 0.22 \times10^{-13}$ 
erg cm$^{-2}$ s$^{-1}$. The  K-corrected luminosity in the 2--10 keV  
energy band is  $L_{2-10 keV} = 6.32_{-0.73}^{+0.76} \times10^{44}$ 
h$^{-2}_{50}$  erg s$^{-1}$ while the  bolometric luminosity is 
$L_{BOL,X} = 1.35^{+0.08}_{-0.21} \times10^{45}$ h$_{50}^{-2}$ erg s$^{-1}$. 
The luminosity in the 0.5--2.0 keV obtained with XMM-{\it Newton} is
$L_{0.5-2.0 keV} = 4.04^{+0.47}_{-0.58} \times10^{44}$ h$^{-2}_{50}$ 
erg s$^{-1}$, comparable with the {\it ROSAT} PSPC luminosity 
of  $L_{X,{\it ROSAT}} = (2.91 \pm 0.63) \times10^{44}$ h$^{-2}_{50}$ 
erg s$^{-1}$. In the concordance cosmology the K-corrected
luminosity in the 2--10 keV energy band becomes 
$L_{2-10 keV} = 5.48^{+0.56}_{-0.74} \times10^{44}$ h$_{70}^{-2}$  
erg s$^{-1}$ and  
the bolometric luminosity becomes $L_{BOL,X} = 1.17^{+0.13}_{-0.18} 
\times10^{45}$ h$_{70}^{-2}$ erg s$^{-1}$.

As can be seen from   Fig.~\ref{Fig2} there are point-like sources both
to the north-east and the south-west which are included 
in the 100\arcsec ~radius even if it is unclear if they belong to the cluster.
The data were re-analyzed excluding the 
point sources from the measure of the cluster X-ray temperature.
Freezing as before the Galactic absorption to the  neutral hydrogen column 
density  N$_{H}=5.24\times10^{20}$ atoms cm$^{-2}$, the redshift to 
z$=$0.8163 and leaving the abundance free to vary, a value 
of T$=$4.9$^{+1.6}_{-0.9}$ keV  is determined for the cluster gas once
the point sources are excluded.  The fit has a reduced $\chi^{2}$ of 
0.91 for 85 degrees of freedom.  This value for the temperature is perfectly
consistent with  the temperature previously  obtained 
(T$=$4.7$^{+1.2}_{-0.7}$  keV) within the uncertainties.

\subsection{RXJ\,1821.9$+$6818, or NEP\,5330: the QSO at z$=$1.692 
in the  XMM-{\it Newton}  field}

There are at least 20 point sources in the XMM-{\it Newton} image. 
We  discuss  here  only one point source, RXJ\,1821.9$+$6818, which appears 
in the NEP survey catalog  \citep{gio03} as  NEP\,5330, and which is
not visible in Fig. 2 since it is outside the field of view.
The source  was identified with an AGN1 (or type 1 AGN, a classification 
which includes either  Seyfert 1 and QSO objects) at redshift 
z$=$1.692$\pm$0.008.  The object was 
spectroscopically observed at the UH 2.2m telescope as part of the 
identification program of the NEP survey sources on June 18, 1998. 
The telescope was equipped  with the Wide Field Grism Spectrograph
and the Tek2048 CCD. 
We used the  420 l mm$^{-1}$ red grism and a 300$\mu$ slit (2.4\arcsec) 
which provided at the f/10 focus a pixel  scale in 
spectroscopic mode of 3.6 \AA\ pix$^{-1}$, a spectral
resolution of $\sim19$ \AA\ FWHM, and a wavelength coverage of
approximately 3900\AA--9500\AA.

The classification with an AGN1 has been done on the basis of the
equivalent width of the emission lines (W$_{\lambda} \sim$30 \AA) and 
on the FWHM ($\sim$4000 km s$^{-1}$) of the broad permitted emission 
lines, indicating a QSO type object (see \citeauthor{gio03} 2003 for 
more details on the optical classification). 
An optical spectrum is shown in Fig.~\ref{Fig4} where the dashed lines 
indicate the positions of the emission lines at the QSO redshift. 
Wavelenghts of atmosphric absorption bands are also indicated. 

To examine the X-ray source associated to the QSO, a circle of radius 
25\arcsec ~was  centered at $\alpha_{2000}= 18^{h}~21^{m}~58.3^{s},  
~\delta_{2000} = +68\deg~18\arcmin ~41.2\arcsec$.
There are a total of 609$\pm$25 net counts in the (0.4--3.8 keV) energy 
range in the MOS1 detector, 632$\pm$26 in (0.5--4.6 keV) in the  MOS2, and 
874$\pm$30 net counts in  (0.5--6.2 keV) in the pn  detector (see Table 2).
The data were fitted in XSPEC  with an absorbed power law model given the 
nature  of the source. The relative normalization of the three 
instruments was left free. The X-ray flux reported below is computed 
using the MOS1 detector  and calibration.  The normalization of the  
MOS2  is  2\% lower than  that of the MOS1 whereas the normalization of 
the pn is 3\% lower. 
The three instrument normalizations are in better agreement than in the 
analysis of the cluster source implying that the high normalization 
discrepancy (24\%) found previously for the pn with respect to the two 
MOS detectors is most probably due to the fact that the cluster falls in 
one of the gap of the pn.

Both the absorption and the photon index ($\Gamma$) of the power law model
were free to vary. The best fit $\Gamma$ is 1.67$\pm$0.12 and the best fit 
absorption is N$_{H}=4.1_{-3.0}^{+3.1}\times10^{20}$ atoms cm$^{-2}$, 
consistent with the  Galactic value of N$_{H}=5.24\times10^{20}$ 
atoms cm$^{-2}$. The binned X-ray spectrum and best fit folded model are 
shown in  Fig.~\ref{Fig5} where bins were chosen so as to have each resulting 
energy  channel significant at the signal-to-noise  ratio of at least 5. 
The fit has a  reduced  $\chi^{2}$ of 0.92 for 61 degrees of freedom. 
The unabsorbed flux of the QSO in the 2--10 keV energy band is
$F_{2-10 keV} = (3.66\pm0.39) \times10^{-13}$ erg cm$^{-2}$ s$^{-1}$
(see Table 3). The  K-corrected luminosity in the 2--10 keV energy band is  
$L_{2-10 keV} = 5.0_{-0.7}^{+0.5} \times10^{45}$ h$^{-2}_{50}$  erg 
s$^{-1}$. In the concordance cosmology the K-corrected luminosity in the 
2--10 keV energy band becomes  $L_{2-10 keV} = (5.3^{+0.6}_{-0.5}) 
\times10^{45}$ h$_{70}^{-2}$   erg s$^{-1}$.
Note that the luminosity of RXJ\,1821.9$+$6818 as measured with 
XMM-{\it Newton} in 2002 is approximately 2.5 times higher than that 
measured with ROSAT in 1990 (see  \citeauthor{mul04} 2004 for revised 
fluxes and luminosities for NEP AGN).  
Such a difference is not surprising given the time variability of AGN.

RXJ\,1821.9$+$6818 was also observed with the Very Large Array at 1.5 GHz 
by \cite{kol94}  who mapped a  29.3 deg$^{2}$ region 
surrounding the north ecliptic pole to support the deepest portion of the 
{\it ROSAT} all-sky soft X-ray survey. RXJ\,1821.9$+$6818 was detected by the
the VLA-NEP survey with a flux density of 182 mJy at  1.5 GHz.

\section{Discussion and Conclusions}

The main result of this paper is the presentation of the optical and X-ray 
data, namely the spectroscopic data acquired through Keck-I, CFH and
UH 2.2m telescopes, and the X-ray  temperature obtained with the detectors 
onboard the XMM-{\it Newton} satellite. 
We will focus the following discussion on the implications of measured 
quantities such as the emission weighted  temperature and velocity 
dispersion for RXJ\,1821.6$+$6827. As with other  X-ray selected,  distant 
clusters \rxa does not have a spherically symmetric morphology in X-rays 
or in optical.  Its X-ray morphology is elongated and very similar to 
other $\sim$0.8 clusters like \rxb \citep{gio99},  another cluster 
found in the course of the NEP identification program but which did not meet 
the 4$\sigma$ selection  criterium for sources \citep{gio03}, or to the better 
known  Medium Survey cluster \ms  \citep{gio04}. Differently from \rxb
and \ms  that both show high velocity dispersions 
\citep[$\sigma_{V}=1522^{+215}_{-150}$ km s$^{-1}$ and 
$\sigma_{V}=1150\pm97$   km s$^{-1}$, respectively:][]
{gio99, vd00}, \rxa has a  velocity dispersion of 
$\sigma_{V}=775^{+182}_{-113}$ km s$^{-1}$, rather typical of normal relaxed 
clusters.  We are also dealing with a cool  cluster with a temperature   
(T$=$4.7 keV) a little lower than the  temperature of \rxb  \citep[which 
{\it Chandra}  measured to be  T$=$6.6$\pm$0.8 keV;][]{vik02} and even lower 
than the X-ray temperature  of \ms  \citep[which XMM-{\it Newton}  measured 
to be T$=$7.2$^{+0.7}_{-0.6}$ keV;][] {gio04}.

A large number of authors (see Table 5 in Girardi et al. 1996,
or Table 2 in Wu et al. 1998 for an exhaustive list of
papers on the subject)  have attempted to determine the
$\sigma_{V}-$T$_{X}$  using different cluster samples in order to test the 
dynamical properties of clusters \citep[see among others][]{gir98, don99, 
tran99, gio99, hms99, rbn02, ye03, lmp04}. 
This relationship is physically meaningful 
since both the velocity dispersion of the galaxies and the temperature of 
the intracluster medium provide a measure of the overall mass of the system. 
If we characterize  the  $\sigma_{V}-$T$_{X}$ relationship  as
$$\beta = { \mu m_p \sigma_{V}^{2} \over k T_{gas} }$$
(where $\mu$ is the mean molecular weight of the gas, m$_{p}$ is
the proton mass and $k$ is the Boltzmann's constant) and if we
assume energy equipartition between the galaxies and the gas in the
cluster ($\beta$$=$1), we  obtain a  velocity dispersion for \rxa consistent
with the measured one. We derive $\sigma_{V}=848^{+101}_{-67}$ km s$^{-1}$ 
from  the above $\sigma_{V}-$T$_{X}$ relation vs the measured value of
$\sigma_{V}=775^{+182}_{-113}$ km s$^{-1}$. 

\citet{gir96} have derived a best fit
relation between the velocity dispersion and the X-ray temperature,
with more than 30\% reduced scatter with respect to other workers
\citep{es91, lb93, bmm95, wfx98}.
\citet{gir96} have taken into account distortions in the velocity
fields, asphericity of the cluster or presence of substructures to
derive their best fit relation:
log($\sigma_{V}$) $=$ (2.53$\pm$0.04) $+$ (0.61$\pm$0.05) log(T). 
If we insert the temperature of \rxa ~in the above relation, the resulting
velocity dispersion value, $\sigma_{V}$ = 874$^{+128}_{-83}$ km s$^{-1}$, 
is consistent, even if on the high side, with the measured velocity dispersion.


The temperature of \rxa is a little lower than predicted from its 
X-ray luminosity for local  L$_{X}-$T$_{X}$ relation.
Understanding the evolution of the  L$_{X}-T_{X}$ relation is important not 
only to understand the  physics behind the formation of galaxy clusters, but
also because it  provides a link between observations of clusters and
derivation  of  cosmological parameters. 
There is an extensive literature on the correlation between these two
basic and measurable quantities based on {\it ASCA}, {\it ROSAT} PSPC and 
more recently {\it Chandra}  and XMM-{\it Newton} data \citep[see
among other][]{dav93, fab94, ms97, mar98, sbo98, ae99, fai00, bor01, nsh02,
vik02, lum04}. The L$_{X}-T_{X}$ relation has been well studied at low 
redshift. Characterizing the L$_{X}$--T$_{X}$
relation as $$ L_{X} = L_{6}({T  \over 6~keV})^{\alpha} $$
investigators have found for $\alpha$ values ranging from 2.33 to 2.88,
all consistent within the errors.
There is no consensus yet on the evolution  of the L$_{X}-T_{X}$ relation 
with redshift \citep[see among others:][]{bor01, nsh02, vik02, lum04} 
probably due to the lack of  large  samples of  galaxy clusters at
cosmologically significant redshift. \citet{vik02} have detected  for the 
first time  an evolutionary  trend in the  L$_{X}$--T$_{X}$ relation using 
{\it Chandra} data for distant  clusters.  More recently \citet{lum04} have 
also found an evolutionary trend of the relation using XMM-{\it Newton} 
data, albeit using a sample of only 8 clusters in a restricted redshift 
range (0.45$<z<$0.62). On the other hand \cite{ett04} detect only hints of 
negative evolution in the L$_{X}$--T$_{X}$ using a larger sample
of 28 clusters observed by Chandra and in a larger redshift range 
(0.4$<z<$1.3). We assume the parametrization by \citet{ae99} who
analyzed a sample of 24 low-z clusters with accurate ASCA temperature 
measurements and absence of strong cooling flows. Comparing
the bolometric luminosity of \rxa ~with the \citet{ae99} best fit relation, 
log(L$_{BOL, X}$) $=$ (2.88$\pm$0.15) log(T/6keV) $+$ (45.06$\pm$0.03),
one would expect for \rxa a temperature of 6.34$^{+0.13}_{-0.35}$ keV that 
is a little higher, even within the uncertainties, than the best-fit 
XMM-{\it Newton}  temperature, implying that this cluster  may fit better 
in an evolving  L$_{X}-$T$_{X}$  relation.
Given the XMM-{\it Newton} results, \rxa does not show
extreme  X-ray luminosity or temperature values.

We can also estimate the mass of \rxa using the X-ray temperature value.
With the assumptions that the mean density in the virialized region is 
$\sim$200 times the critical density at the redshift of the cluster and that 
the cluster is  isothermal \citep{emn96, don98},
we can use the  scaling law method as illustrated in \citet{ae99}. 
From the simulations of Evrard, Metzler \& Navarro (1996) for the 
mass-temperature relation one can estimate the virial mass within a radius
$r_{200}(T) = 1.85 ({T\over 10keV})^{1/2} (1+z)^{-3/2} h^{-1} Mpc $ 
by using the  equation   
$$ M_{vir} \approx (1.45\times10^{15} h^{-1} M_{\sun})(1+z)^{-3/2}
({kT_{X} \over 10 keV})^{1.5} $$
From the XMM-{\it Newton} temperature the virial mass is 
approximately M$_{vir} = (1.9^{+0.8}_{-0.4})\times10^{14}~h^{-1}_{50}$ 
M$_{\sun}$ ~within  $r_{200}=1.04 ~h^{-1}_{50}$ Mpc, where the uncertainties 
on the mass  reflect the  uncertainties on the temperature. 


\section{Summary}

We have presented observations for the z$=$0.8163 galaxy  cluster \rxa  
performed with the instruments on board the XMM-{\it Newton} satellite,
and with the CFHT and Keck-I telescopes. 
The main result of the paper is to present new optical and
X-ray data for the most distant cluster in the NEP survey.
Both the temperature and the metal abundance have been left as  
free parameters in the fitting of the X-ray data with a model. Thus
freezing the hydrogen column density to the Galactic value of 
$5.24\times10^{20}$ atoms cm$^{-2}$ and the redshift to the measured
z$=$0.8163, we obtain a best fit temperature with a  thermal MEKAL model 
of T$=$4.7$^{+1.2}_{-0.7}$ keV, while the abundance is unconstrained
and can vary in the range 0.28--1.42 Z$_{\sun}$.
This X-ray temperature is a little lower than  predictions from the cluster
X-ray bolometric luminosity  $L_{BOL,X} = 1.35^{+0.08}_{-0.21} \times10^{45}
h_{50}^{-2}$ erg s$^{-1}$ ($L_{BOL,X} = 1.17^{+0.13}_{-0.18} \times10^{45}
h_{70}^{-2}$ erg s$^{-1}$ in the concordance cosmology) from the L$_{X}-T_{X}$
relation of local clusters published in the literature.

From 18 accepted cluster members we obtain an average redshift  
$<z>=0.8163\pm0.0011$ and a dispersion along the line
of sight  $\sigma_{V}=775^{+182}_{-113}$  km s$^{-1}$.
This measured value is consistent with the velocity dispersion 
expected from  the  $\sigma_{V}-T_{X}$  relationship obtained from 
different cluster samples.  From the M -- T relation we  estimate a 
virial mass for \rxa within the $r_{200}=1.04 ~h^{-1}_{50}$ Mpc on the 
assumptions  that the mean density in the virialized region is 
$\sim$200 times the critical density at the redshift of the cluster and that 
the cluster is  isothermal. The estimated virial mass has a value of
M$_{vir}=1.9^{+0.8}_{-0.4} \times10^{14}~h^{-1}_{50}$ M$_{\sun}$ which
shows that \rxa  is a massive cluster at high redshift.

The point source to the south, RXJ\,1821.9$+$6818, identified in the
NEP survey as a QSO at z$=$1.692$\pm$0.008, is well fitted by an 
absorbed power  law with N$_{H}=4.1_{-3.0}^{+3.1}\times10^{20}$ atoms 
cm$^{-2}$  and a photon index $\Gamma=1.67\pm$0.12 which is typical 
for this class of objects.


\begin{acknowledgements}

IMG would like to thank the hospitality of the Institute for Astronomy of the
University of Hawai$'$i where this paper was written. She also notes 
that  this work was done in spite of  the continued efforts by the 
Italian government to dismantle publicly-funded fundamental research.
An anonymous referee made several comments which improved the manuscript. 
Partial financial support for this work came from the Italian Space 
Agency ASI (Agenzia Spaziale Italiana) through grant ASI I/R/037/00.

\end{acknowledgements}

\bibliographystyle{aa}
\bibliography{myrefs}

\begin{thebibliography}{59}
\expandafter\ifx\csname natexlab\endcsname\relax\def\natexlab#1{#1}\fi

\bibitem[{{Anders} \& {Grevesse}(1989)}]{ag89}
{Anders}, E. \& {Grevesse}, N. 1989, \gca, 53, 197

\bibitem[{{Arnaud} \& {Evrard}(1999)}]{ae99}
{Arnaud}, M. \& {Evrard}, A.~E. 1999, \mnras, 305, 631

\bibitem[{{Beers} {et~al.}(1990){Beers}, {Flynn}, \& {Gebhardt}}]{beers90}
{Beers}, T.~C., {Flynn}, K., \& {Gebhardt}, K. 1990, \aj, 100, 32

\bibitem[{{Bird} \& {Beers}(1993)}]{bb93}
{Bird}, C.~M. \& {Beers}, T.~C. 1993, \aj, 105, 1596

\bibitem[{{Bird} {et~al.}(1995){Bird}, {Mushotzky}, \& {Metzler}}]{bmm95}
{Bird}, C.~M., {Mushotzky}, R.~F., \& {Metzler}, C.~A. 1995, \apj, 453, 40

\bibitem[{{Borgani} {et~al.}(1999){Borgani}, {Rosati}, {Tozzi}, \&
  {Norman}}]{bor99}
{Borgani}, S., {Rosati}, P., {Tozzi}, P., \& {Norman}, C. 1999, \apj, 517, 40

\bibitem[{{Borgani} {et~al.}(2001){Borgani}, {Rosati}, {Tozzi}, {Stanford},
  {Eisenhardt}, {Lidman}, {Holden}, {Della Ceca}, {Norman}, \&
  {Squires}}]{bor01}
{Borgani}, S., {Rosati}, P., {Tozzi}, P., {et~al.} 2001, \apj, 561, 13

\bibitem[{{Danese} {et~al.}(1980){Danese}, {de Zotti}, \& {di Tullio}}]{dan80}
{Danese}, L., {de Zotti}, G., \& {di Tullio}, G. 1980, \aap, 82, 322

\bibitem[{{David} {et~al.}(1993){David}, {Slyz}, {Jones}, {Forman}, {Vrtilek},
  \& {Arnaud}}]{dav93}
{David}, L.~P., {Slyz}, A., {Jones}, C., {et~al.} 1993, \apj, 412, 479

\bibitem[{{Donahue} {et~al.}(1998){Donahue}, {Voit}, {Gioia}, {Lupino},
  {Hughes}, \& {Stocke}}]{don98}
{Donahue}, M., {Voit}, G.~M., {Gioia}, I., {et~al.} 1998, \apj, 502, 550

\bibitem[{{Donahue} {et~al.}(1999){Donahue}, {Voit}, {Scharf}, {Gioia},
  {Mullis}, {Hughes}, \& {Stocke}}]{don99}
{Donahue}, M., {Voit}, G.~M., {Scharf}, C.~A., {et~al.} 1999, \apj, 527, 525

\bibitem[{{Edge} \& {Stewart}(1991)}]{es91}
{Edge}, A.~C. \& {Stewart}, G.~C. 1991, \mnras, 252, 414

\bibitem[{{Eke} {et~al.}(1998){Eke}, {Cole}, {Frenk}, \& {Patrick
  Henry}}]{eke98}
{Eke}, V.~R., {Cole}, S., {Frenk}, C.~S., \& {Patrick Henry}, J. 1998, \mnras,
  298, 1145

\bibitem[{{Elvis} {et~al.}(1994){Elvis}, {Lockman}, \& {Fassnacht}}]{elv94}
{Elvis}, M., {Lockman}, F.~J., \& {Fassnacht}, C. 1994, \apjs, 95, 413

\bibitem[{{Ettori} {et~al.}(2004){Ettori}, {Tozzi}, {Borgani}, \&
  {Rosati}}]{ett04}
{Ettori}, S., {Tozzi}, P., {Borgani}, S., \& {Rosati}, P. 2004, \aap, 417, 13

\bibitem[{{Evrard} {et~al.}(1996){Evrard}, {Metzler}, \& {Navarro}}]{emn96}
{Evrard}, A.~E., {Metzler}, C.~A., \& {Navarro}, J.~F. 1996, \apj, 469, 494

\bibitem[{{Fabian} {et~al.}(1994){Fabian}, {Crawford}, {Edge}, \&
  {Mushotzky}}]{fab94}
{Fabian}, A.~C., {Crawford}, C.~S., {Edge}, A.~C., \& {Mushotzky}, R.~F. 1994,
  \mnras, 267, 779

\bibitem[{{Fairley} {et~al.}(2000){Fairley}, {Jones}, {Scharf}, {Ebeling},
  {Perlman}, {Horner}, {Wegner}, \& {Malkan}}]{fai00}
{Fairley}, B.~W., {Jones}, L.~R., {Scharf}, C., {et~al.} 2000, \mnras, 315, 669

\bibitem[{{Gioia} {et~al.}(2004){Gioia}, {Braito}, {Branchesi}, {Della Ceca},
  {Maccacaro}, \& {Tran}}]{gio04}
{Gioia}, I.~M., {Braito}, V., {Branchesi}, M., {et~al.} 2004, \aap, 419, 517

\bibitem[{{Gioia} {et~al.}(2003){Gioia}, {Henry}, {Mullis}, {B{\" o}hringer},
  {Briel}, {Voges}, \& {Huchra}}]{gio03}
{Gioia}, I.~M., {Henry}, J.~P., {Mullis}, C.~R., {et~al.} 2003, \apjs, 149, 29

\bibitem[{{Gioia} {et~al.}(1999){Gioia}, {Henry}, {Mullis}, {Ebeling}, \&
  {Wolter}}]{gio99}
{Gioia}, I.~M., {Henry}, J.~P., {Mullis}, C.~R., {Ebeling}, H., \& {Wolter}, A.
  1999, \aj, 117, 2608

\bibitem[{{Gioia} {et~al.}(2001){Gioia}, {Henry}, {Mullis}, {Voges}, {Briel},
  {B{\" o}hringer}, \& {Huchra}}]{gio01}
{Gioia}, I.~M., {Henry}, J.~P., {Mullis}, C.~R., {et~al.} 2001, \apjl, 553,
  L105

\bibitem[{{Girardi} {et~al.}(1996){Girardi}, {Fadda}, {Giuricin},
  {Mardirossian}, {Mezzetti}, \& {Biviano}}]{gir96}
{Girardi}, M., {Fadda}, D., {Giuricin}, G., {et~al.} 1996, \apj, 457, 61

\bibitem[{{Girardi} {et~al.}(1998){Girardi}, {Giuricin}, {Mardirossian},
  {Mezzetti}, \& {Boschin}}]{gir98}
{Girardi}, M., {Giuricin}, G., {Mardirossian}, F., {Mezzetti}, M., \&
  {Boschin}, W. 1998, \apj, 505, 74

\bibitem[{{Grevesse} \& {Sauval}(1998)}]{gs98}
{Grevesse}, N. \& {Sauval}, A.~J. 1998, Space Science Reviews, 85, 161

\bibitem[{{Henry}(2000)}]{hen00}
{Henry}, J.~P. 2000, \apj, 534, 565

\bibitem[{{Henry}(2004)}]{hen04}
{Henry}, J.~P. 2004, ArXiv Astrophysics e-prints

\bibitem[{{Henry} \& {Arnaud}(1991)}]{ha91}
{Henry}, J.~P. \& {Arnaud}, K.~A. 1991, \apj, 372, 410

\bibitem[{{Henry} {et~al.}(2001){Henry}, {Gioia}, {Mullis}, {Voges}, {Briel},
  {B{\" o}hringer}, \& {Huchra}}]{hen01}
{Henry}, J.~P., {Gioia}, I.~M., {Mullis}, C.~R., {et~al.} 2001, \apjl, 553,
  L109

\bibitem[{{Horner} {et~al.}(1999){Horner}, {Mushotzky}, \& {Scharf}}]{hms99}
{Horner}, D.~J., {Mushotzky}, R.~F., \& {Scharf}, C.~A. 1999, \apj, 520, 78

\bibitem[{{Jansen} {et~al.}(2001){Jansen}, {Lumb}, {Altieri}, {Clavel}, {Ehle},
  {Erd}, {Gabriel}, {Guainazzi}, {Gondoin}, {Much}, {Munoz}, {Santos},
  {Schartel}, {Texier}, \& {Vacanti}}]{jan01}
{Jansen}, F., {Lumb}, D., {Altieri}, B., {et~al.} 2001, \aap, 365, L1

\bibitem[{{Kollgaard} {et~al.}(1994){Kollgaard}, {Brinkmann}, {Chester},
  {Feigelson}, {Hertz}, {Reich}, \& {Wielebinski}}]{kol94}
{Kollgaard}, R.~I., {Brinkmann}, W., {Chester}, M.~M., {et~al.} 1994, \apjs,
  93, 145

\bibitem[{{Le Fevre} {et~al.}(1995){Le Fevre}, {Crampton}, {Lilly}, {Hammer},
  \& {Tresse}}]{olf95}
{Le Fevre}, O., {Crampton}, D., {Lilly}, S.~J., {Hammer}, F., \& {Tresse}, L.
  1995, \apj, 455, 60

\bibitem[{{Lubin} \& {Bahcall}(1993)}]{lb93}
{Lubin}, L.~M. \& {Bahcall}, N.~A. 1993, \apjl, 415, L17

\bibitem[{{Lubin} {et~al.}(2004){Lubin}, {Mulchaey}, \& {Postman}}]{lmp04}
{Lubin}, L.~M., {Mulchaey}, J.~S., \& {Postman}, M. 2004, \apjl, 601, L9

\bibitem[{{Lumb} {et~al.}(2004){Lumb}, {Bartlett}, {Romer}, {Blanchard},
  {Burke}, {Collins}, {Nichol}, {Giard}, {Marty}, {Nevalainen}, {Sadat}, \&
  {Vauclair}}]{lum04}
{Lumb}, D.~H., {Bartlett}, J.~G., {Romer}, A.~K., {et~al.} 2004, ArXiv
  Astrophysics e-prints

\bibitem[{{Markevitch}(1998)}]{mar98}
{Markevitch}, M. 1998, \apj, 504, 27

\bibitem[{{Mewe} {et~al.}(1985){Mewe}, {Gronenschild}, \& {van den
  Oord}}]{mew85}
{Mewe}, R., {Gronenschild}, E.~H.~B.~M., \& {van den Oord}, G.~H.~J. 1985,
  \aaps, 62, 197

\bibitem[{{Mullis} {et~al.}(2001){Mullis}, {Henry}, {Gioia}, {B{\" o}hringer},
  {Briel}, {Voges}, \& {Huchra}}]{mul01}
{Mullis}, C.~R., {Henry}, J.~P., {Gioia}, I.~M., {et~al.} 2001, \apjl, 553,
  L115

\bibitem[{{Mullis} {et~al.}(2004){Mullis}, {Henry}, {Gioia}, {B{\" o}hringer},
  {Briel}, {Voges}, \& {Huchra}}]{mul04}
{Mullis}, C.~R., {Henry}, J.~P., {Gioia}, I.~M., {et~al.} 2004, ApJ, submitted

\bibitem[{{Mushotzky} \& {Scharf}(1997)}]{ms97}
{Mushotzky}, R.~F. \& {Scharf}, C.~A. 1997, \apjl, 482, L13

\bibitem[{{Novicki} {et~al.}(2002){Novicki}, {Sornig}, \& {Henry}}]{nsh02}
{Novicki}, M.~C., {Sornig}, M., \& {Henry}, J.~P. 2002, \aj, 124, 2413

\bibitem[{{Oke} {et~al.}(1995){Oke}, {Cohen}, {Carr}, {Cromer}, {Dingizian},
  {Harris}, {Labrecque}, {Lucinio}, {Schaal}, {Epps}, \& {Miller}}]{oke95}
{Oke}, J.~B., {Cohen}, J.~G., {Carr}, M., {et~al.} 1995, \pasp, 107, 375

\bibitem[{{Oukbir} \& {Blanchard}(1992)}]{ob92}
{Oukbir}, J. \& {Blanchard}, A. 1992, \aap, 262, L21

\bibitem[{{Oukbir} \& {Blanchard}(1997)}]{ob97}
{Oukbir}, J. \& {Blanchard}, A. 1997, \aap, 317, 1

\bibitem[{{Peacock}(1999)}]{pea99}
{Peacock}, J.~A. 1999, {Cosmological physics} (Cosmological physics.~
  Publisher: Cambridge, UK: Cambridge University Press, 1999.~ISBN: 0521422701)

\bibitem[{{Peebles}(1993)}]{peeb93}
{Peebles}, P.~J.~E. 1993, {Principles of physical cosmology} (Princeton Series
  in Physics, Princeton, NJ: Princeton University Press, |c1993)

\bibitem[{{Rosati} {et~al.}(2002){Rosati}, {Borgani}, \& {Norman}}]{rbn02}
{Rosati}, P., {Borgani}, S., \& {Norman}, C. 2002, \araa, 40, 539

\bibitem[{{Rosati} {et~al.}(2004){Rosati}, {Tozzi}, {Ettori}, {Mainieri},
  {Demarco}, {Stanford}, {Lidman}, {Nonino}, {Borgani}, {Della Ceca},
  {Eisenhardt}, {Holden}, \& {Norman}}]{ros04}
{Rosati}, P., {Tozzi}, P., {Ettori}, S., {et~al.} 2004, \aj, 127, 230

\bibitem[{{Sadat} {et~al.}(1998){Sadat}, {Blanchard}, \& {Oukbir}}]{sbo98}
{Sadat}, R., {Blanchard}, A., \& {Oukbir}, J. 1998, \aap, 329, 21

\bibitem[{{Str{\" u}der} {et~al.}(2001){Str{\" u}der}, {Briel}, {Dennerl},
  {Hartmann}, {Kendziorra}, {Meidinger}, {Pfeffermann}, {Reppin}, {Aschenbach},
  {Bornemann}, {Br{\" a}uninger}, {Burkert}, {Elender}, {Freyberg}, {Haberl},
  {Hartner}, {Heuschmann}, {Hippmann}, {Kastelic}, {Kemmer}, {Kettenring},
  {Kink}, {Krause}, {M{\" u}ller}, {Oppitz}, {Pietsch}, {Popp}, {Predehl},
  {Read}, {Stephan}, {St{\" o}tter}, {Tr{\" u}mper}, {Holl}, {Kemmer},
  {Soltau}, {St{\" o}tter}, {Weber}, {Weichert}, {von Zanthier},
  {Carathanassis}, {Lutz}, {Richter}, {Solc}, {B{\" o}ttcher}, {Kuster},
  {Staubert}, {Abbey}, {Holland}, {Turner}, {Balasini}, {Bignami}, {La
  Palombara}, {Villa}, {Buttler}, {Gianini}, {Lain{\' e}}, {Lumb}, \&
  {Dhez}}]{sru01}
{Str{\" u}der}, L., {Briel}, U., {Dennerl}, K., {et~al.} 2001, \aap, 365, L18

\bibitem[{{Tran} {et~al.}(1999){Tran}, {Kelson}, {van Dokkum}, {Franx},
  {Illingworth}, \& {Magee}}]{tran99}
{Tran}, K.~H., {Kelson}, D.~D., {van Dokkum}, P., {et~al.} 1999, \apj, 522, 39

\bibitem[{{Turner} {et~al.}(2001){Turner}, {Abbey}, {Arnaud}, {Balasini},
  {Barbera}, {Belsole}, {Bennie}, {Bernard}, {Bignami}, {Boer}, {Briel},
  {Butler}, {Cara}, {Chabaud}, {Cole}, {Collura}, {Conte}, {Cros}, {Denby},
  {Dhez}, {Di Coco}, {Dowson}, {Ferrando}, {Ghizzardi}, {Gianotti}, {Goodall},
  {Gretton}, {Griffiths}, {Hainaut}, {Hochedez}, {Holland}, {Jourdain},
  {Kendziorra}, {Lagostina}, {Laine}, {La Palombara}, {Lortholary}, {Lumb},
  {Marty}, {Molendi}, {Pigot}, {Poindron}, {Pounds}, {Reeves}, {Reppin},
  {Rothenflug}, {Salvetat}, {Sauvageot}, {Schmitt}, {Sembay}, {Short},
  {Spragg}, {Stephen}, {Str{\" u}der}, {Tiengo}, {Trifoglio}, {Tr{\" u}mper},
  {Vercellone}, {Vigroux}, {Villa}, {Ward}, {Whitehead}, \& {Zonca}}]{tur01}
{Turner}, M.~J.~L., {Abbey}, A., {Arnaud}, M., {et~al.} 2001, \aap, 365, L27

\bibitem[{{van Dokkum} {et~al.}(2000){van Dokkum}, {Franx}, {Fabricant},
  {Illingworth}, \& {Kelson}}]{vd00}
{van Dokkum}, P.~G., {Franx}, M., {Fabricant}, D., {Illingworth}, G.~D., \&
  {Kelson}, D.~D. 2000, \apj, 541, 95

\bibitem[{{Vikhlinin} {et~al.}(2002){Vikhlinin}, {VanSpeybroeck}, {Markevitch},
  {Forman}, \& {Grego}}]{vik02}
{Vikhlinin}, A., {VanSpeybroeck}, L., {Markevitch}, M., {Forman}, W.~R., \&
  {Grego}, L. 2002, \apjl, 578, L107

\bibitem[{{Vikhlinin} {et~al.}(2003){Vikhlinin}, {Voevodkin}, {Mullis},
  {VanSpeybroeck}, {Quintana}, {McNamara}, {Gioia}, {Hornstrup}, {Henry},
  {Forman}, \& {Jones}}]{vik03}
{Vikhlinin}, A., {Voevodkin}, A., {Mullis}, C.~R., {et~al.} 2003, \apj, 590, 15

\bibitem[{{Voges} {et~al.}(2001){Voges}, {Henry}, {Briel}, {B{\" o}hringer},
  {Mullis}, {Gioia}, \& {Huchra}}]{vog01}
{Voges}, W., {Henry}, J.~P., {Briel}, U.~G., {et~al.} 2001, \apjl, 553, L119

\bibitem[{{Wu} {et~al.}(1998){Wu}, {Fang}, \& {Xu}}]{wfx98}
{Wu}, X., {Fang}, L., \& {Xu}, W. 1998, \aap, 338, 813

\bibitem[{{Yee} \& {Ellingson}(2003)}]{ye03}
{Yee}, H.~K.~C. \& {Ellingson}, E. 2003, \apj, 585, 215

\end{thebibliography}

\newpage

\begin{table*}
  \begin{center}
    \caption{Spectroscopic data for the galaxies in the  NEP cluster 
     RXJ\,1821.6$+$6827}

\begin{tabular}{rllcrcl}

& & & & & &\\

\hline
\hline
           &             &       &   &  &   & \\
ID & RA (J2000) & DEC (J2000) & cz  &  $\Delta$\,cz  & z & Spectral features \\
\#& ~~h ~~m ~~s& ~~~\deg ~~~\arcmin ~~~\arcsec& km s$^{-1}$ & km s$^{-1}$ & & \\

\hline

& & & & & & \\
1 & 18 21 16.6&$+$68 30 25& 244367& 120&0.8151& H$+$K, G band, H$\delta$, CN \\
2  & 18 21 21.5  & $+$68 29 57 & 246585 & 240 & 0.8225 & H$+$K, CN \\
3  & 18 21 26.2  & $+$68 29 31 & 246016 & 170 & 0.8206 & H$+$K, G band, CN \\
4  & 18 21 30.8  & $+$68 29 29 & 245856 & 180 & 0.8204 & H$+$K, G band, CN \\
5  & 18 21 27.9  & $+$68 28 43 & 246705 & 480 & 0.8229 & H$+$K, G band, CN \\
6  & 18 21 33.8  & $+$68 28 40 & 246885 & 210 & 0.8235 & H$+$K \\
7  & 18 21 32.6  & $+$68 27 54 & 244847 & 180 & 0.8167 & H$+$K, G band, cD ?\\
8  & 18 21 34.9  & $+$68 27 53 & 242808 &  90 & 0.8099 & H$+$K, CN \\
9  & 18 21 37.2  & $+$68 27 51 & 244428 & 389 & 0.8153 & H$+$K, G band, CN \\
10 & 18 21 33.6  & $+$68 27 50 & 243317 & 389 & 0.8161 & H$+$K, G band \\
11 & 18 21 35.2  & $+$68 27 34 & 243288 &  60 & 0.8115 & H$+$K, CN \\
12 & 18 21 36.6  & $+$68 27 21 & 244996 & 120 & 0.8172 & H$+$K \\
13 & 18 21 37.5  & $+$68 27 07 & 246765 & 150 & 0.8231 & H$+$K \\
14 & 18 21 37.9  & $+$68 27 00 & 244607 &  60 & 0.8159 & H$+$K \\
15 & 18 21 18.2  & $+$68 27 00 & 238102 & 840 & 0.7942 & [OII] \\
16 & 18 21 47.7  & $+$68 27 06 & 242508 &  30 & 0.8089 & H$+$K \\
17 & 18 21 51.1  & $+$68 27 17 & 241639 & 540 & 0.8060 & CaII Break \\
18 & 18 21 57.7  & $+$68 27 25 & 243378 & 540 & 0.8118 & CaII Break \\
19 & 18 21 50.8  & $+$68 26 09 & 244397 & 840 & 0.8152 & CaII Break, [OII] \\
20 & 18 21 47.8  & $+$68 25 18 & 243498 &  90 & 0.8122 & H$+$K \\

& & & & & & \\

\hline
\hline

\end{tabular}
\end{center}

\end{table*}
\label{tab1}


\begin{table*}
\begin{center}

\caption{X-ray properties of RXJ\,1821.6$+$6827 (NEP\,5281) and 
RXJ\,1821.9$+$6818 (NEP\,5330)}

\begin{tabular}{llllrcccl}

& & & & & & & & \\

\hline
\hline
           &             &       &   &            &    &    &  & \\
RA (J2000) & DEC (J2000) & Instr & R & Net Cts & T  & Z (Z$_{\sun}$)  & N$_{H}$ & Region \\
 ~~h ~~m ~~s & ~~~\deg ~~~\arcmin ~~~\arcsec &  & \arcsec & & keV & $\Gamma$ & $\times10^{20}$ at/cm$^{2}$& \\

\hline

& & & & & & & & \\

18~21~32.5 & +68~27~45.0 & &  & & 4.7$^{+1.2}_{-0.7}$ & 0.77$^{+0.65}_{-0.49}$ & frozen$^{\mathrm{a}}$ &  cluster \\
   & & MOS1 & 100 & 708$\pm$35 & &  &  & \\
   & & MOS2 & 100 & 671$\pm$36 & &  &  & \\
   & & pn   & 100 & 751$\pm$40 & &  &  & \\
18~21~58.3 & +68~18~41.2 & &  & & & 1.67$\pm$0.12 & 4.1$^{+3.1}_{-3.0}$ & QSO \\
   & & MOS1 & 25 & 609$\pm$25 & &  &  & \\
   & & MOS2 & 25 & 632$\pm$26 & &  &  & \\
   & & pn   & 25 & 874$\pm$30 & &  &  & \\

& & & & & & & \\

\hline
\hline

\end{tabular}
\end{center}
\begin{list}{}{}
\item[$^{\mathrm{a}}$]  N$_{H}$ fixed at the Galactic value of 
$5.24\times10^{20}$ atoms cm$^{-2}$
\end{list}

\end{table*}
\label{tab2}
%

\begin{table*}
\begin{center}

\caption{Fluxes and luminosities for the  galaxy cluster and the QSO}

\begin{tabular}{lllcccc}

& & & & & & \\

\hline
\hline
           &             &       &   &  &   \\
RA (J2000) & DEC (J2000) & F$_{2-10 keV}$ & z & L$_{2-10 keV}$ & L$_{BOL}$ & Region \\
 ~~h ~~m ~~s & ~~~\deg ~~~\arcmin ~~~\arcsec & $10^{-13}$ cgs & & $10^{44}$ h$^{-2}_{50}$ cgs$^{\mathrm{a}}$ & $10^{45}$  h$^{-2}_{50}$ cgs  & \\
 &  & & & $10^{44}$ h$^{-2}_{70}$ cgs$^{\mathrm{a}}$ & $10^{45}$ h$^{-2}_{70}$ cgs  &  \\

\hline

& & & & & &  \\
18~21~32.5 & +68~27~45.0 & 1.24$^{+0.16}_{-0.23}$ & 0.816$\pm$0.001 & 6.32$^{+0.76}_{-0.73}$ & 1.35$^{+0.08}_{-0.21}$ & cluster \\
& & & & & & \\
& & & & 5.48$^{+0.56}_{-0.74}$ & 1.17$^{+0.13}_{-0.18}$ &  \\
& & & & & & \\
18~21~58.3 & +68~18~41.2 & 3.66$\pm$0.39 & 1.692$\pm$0.008 & 50$^{+5}_{-7}$ & & QSO\\
& & & & &  & \\
& & & & 53$^{+6}_{-5}$ & &   \\
& &  & & & & \\

\hline
\hline

\end{tabular}
\end{center}
\begin{list}{}{}
\item[$^{\mathrm{a}}$]  First line refers to the Einstein-de-Sitter 
model with  H$_{0}=50$ km s$^{-1}$ Mpc$^{-1}$,  $\Omega_M=1$, and
$\Omega_\Lambda=0$, while second line refers to the 
cosmological concordance model with H$_{0}=70$ km s$^{-1}$ Mpc$^{-1}$,
 $\Omega_M=0.3$, $\Omega_\Lambda=0.7$
\end{list}

\end{table*}
\label{tab3}

\end{document}